# Toward AI-Native 6G Air Interface: A 3GPP Perspective on Protocol Framework

Xingqin Lin
NVIDIA
Email: xingqinl@nvidia.com

*Abstract*—**Artificial intelligence (AI) is expected to play an important role in the sixth-generation (6G) air interface design, but making the air interface truly AI-native requires more than applying learning algorithms to individual radio functions. The deeper challenge is architectural: once AI influences how the user equipment and network interpret, predict, and adapt radio behavior, the air interface must provide common protocol semantics for coordinating such intelligence across vendors and deployments. This article presents a 3rd generation partnership project (3GPP) oriented perspective on the protocol framework for AI-native 6G air interface. We argue that standardization should preserve implementation freedom by avoiding prescription of model architectures, training methods, or model weights. Instead, 6G should define the protocol framework needed for interoperable AI operation, including how AI-enabled functions are configured, validated, activated, monitored, and safely reverted to conventional operation. Neural receiver assisted reference signal adaptation is used as a case study to concretely show this broader architectural shift.**

## I. INTRODUCTION

Cellular air interface standardization has traditionally been built on a careful balance between interoperability and implementation freedom [1]. The standard defines the externally visible behavior that allows a user equipment (UE) and a network from different vendors to communicate reliably, while leaving substantial room for proprietary innovation inside the receiver, scheduler, and radio resource management algorithms. This separation has been one of the foundations of cellular success: the protocol specifications are common, but the implementation choices behind the specifications remain competitive. Artificial intelligence (AI) does not necessarily disturb this balance [2]. If AI is used only to improve an internal algorithm, the existing model of standardization remains largely intact. A UE may use a neural network to improve receiver processing, or a base station (BS) may use reinforcement learning to improve scheduling decisions, without requiring the peer to know how the algorithm works. In such cases, AI is an implementation technique. The air interface remains unchanged, and interoperability is preserved through the same standardized signals, reports, and procedures.

The challenge becomes more fundamental when AI begins to influence behavior that must be coordinated across the air interface. Future sixth-generation (6G) systems are expected to use learning-based methods not only to improve local processing, but also to interpret radio observations, predict future channel states, and enable more proactive radio control [3]. Once such AI outputs affect how the UE and BS configure, interpret, or adapt radio behavior, the protocol must provide more than conventional resource allocations and measurement reports. It must provide common semantics for what the AI-enabled function is doing, when its output is valid, how reliable it is, and how the system should respond when that reliability degrades. This distinction is central to the meaning of an AI-native air interface. An AI-native air interface includes protocol support for learnable and adaptive radio functions [4], [5]. The goal is not to standardize neural network architectures, training datasets, optimization methods, or model weights. Doing so would constrain innovation and would be poorly matched to the rapid evolution of AI techniques and deployment-specific model design. Instead, the standardization challenge is to define the air interface around AI-enabled behavior so that different implementations can interoperate safely.

Note that the 3rd generation partnership project (3GPP) Release 19 already provides feature-specific mechanisms for AI-enabled beam management, channel state information (CSI) prediction, and positioning [6], [7]. These include UE capability signaling, inference configurations or parameter sets, configuration-specific applicability reporting, and feature-dependent activation and monitoring procedures. Besides, TS 28.105 separately specifies management functions for the AI lifecycle [8]. This article generalizes their interoperability principles into a candidate 6G framework for future AI-enabled functions that change peer-visible radio behavior, such as reference signal adaptation. In particular, it adds a common way to describe versioned operating profiles, task-specific reliability semantics, deterministic fallback, and compatible transmitter-receiver assumptions where required. Throughout this article, established mechanisms are explicitly attributed to 3GPP specifications; the AI task profile abstraction and the illustrative radio resource control (RRC), medium access control (MAC) control element (CE), downlink control information (DCI), and uplink control information (UCI) procedures are proposed candidate directions for future 6G standardization.

The rest of this article is organized as follows. We first discuss how AI-native operation changes the air interface design and introduce protocol-visible objects such as AI task profiles, capability descriptors, representation semantics, validity regions, confidence reporting, and fallback profiles. We then map these objects onto a layered protocol architecture and describe a procedure flow for configuration, validation, activation, monitoring, and recovery. Finally, we use neural receiver assisted reference signal adaptation as a concrete example of how AI can enable deeper physical layer innovation while still preserving the standardization principles that have made cellular systems reliable and multi-vendor interoperable. Figure 1 provides an illustration of the AI-native 6G air interface protocol framework and operation procedure that will be discussed in detail in the rest of this article.

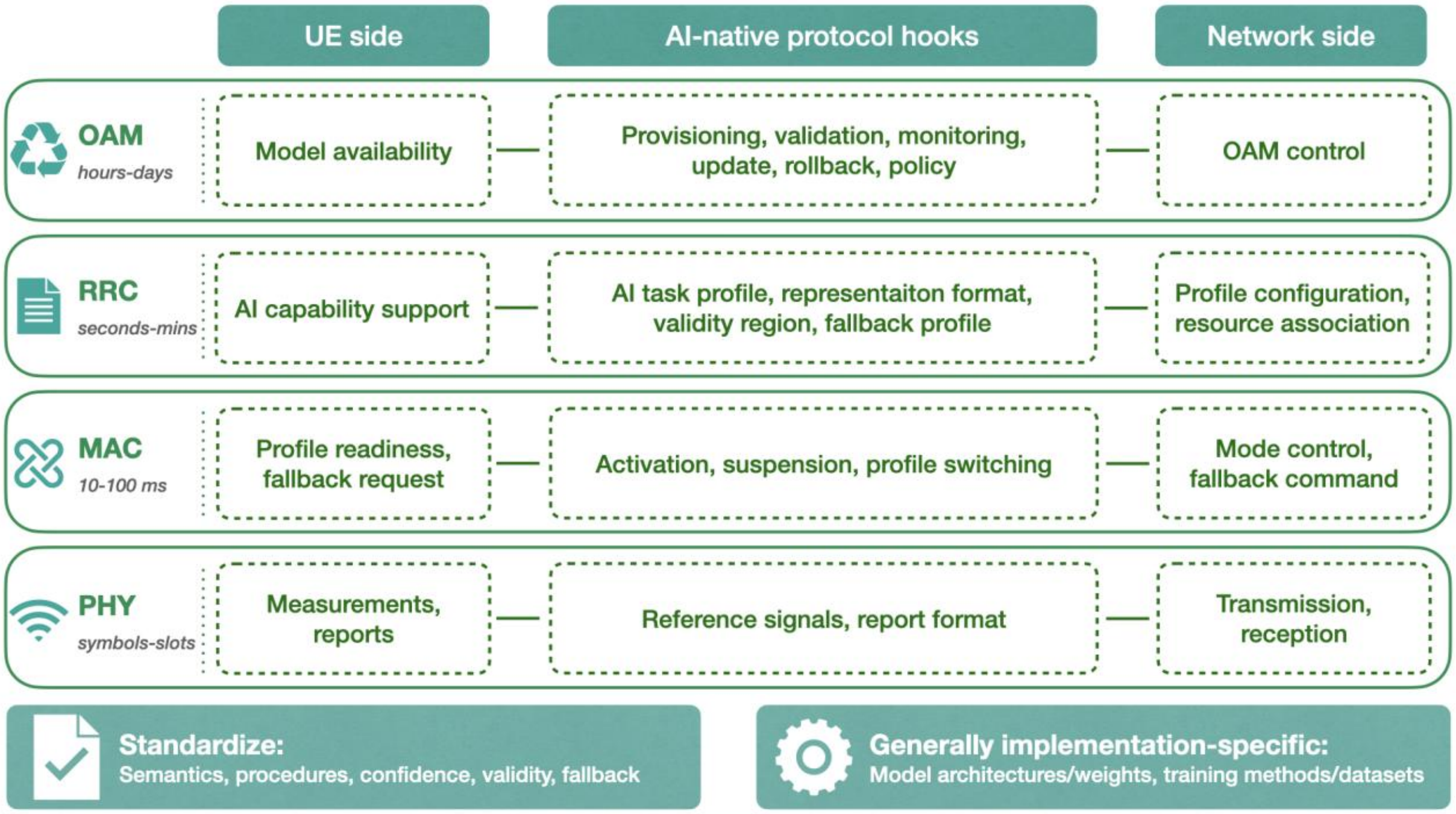

**(a) Protocol framework**

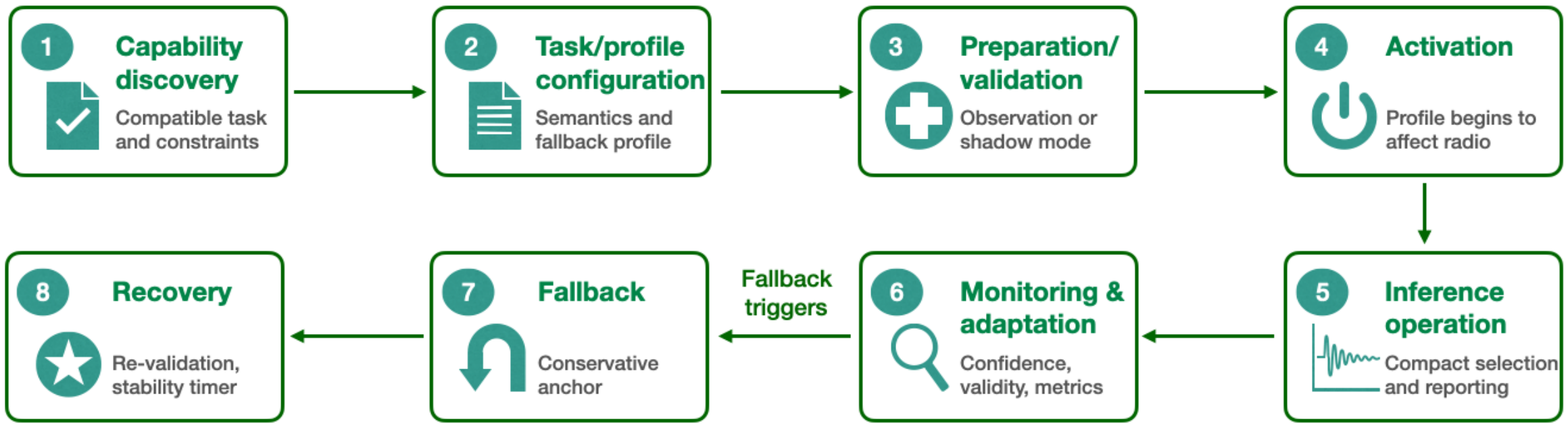

**(b) Procedure flow**

**Figure 1: An illustration of proposed AI-native 6G air interface protocol framework and operation procedure.**

## II. STANDARDIZATION OBJECTS

The air interface specifications define how radio resources are configured, how transmitted signals are interpreted, how measurements are reported, and how the system recovers when radio conditions change. In existing cellular systems, the specifications are largely deterministic from a protocol perspective. Even when the UE and BS use sophisticated proprietary algorithms internally, the meaning of the exchanged information is standardized and stable. AI-native operation changes this because the behavior of a radio function may depend on learned representations, prediction models, deployment context, inference timing, and operating conditions [9]. For example, a conventional CSI report has a standardized interpretation regardless of how the UE computes it. By contrast, an AI-assisted CSI output may represent a compressed channel description, a predicted future channel state, or an uncertainty-aware estimate [10]. These outputs can be useful only if the BS understands their intended meaning and the conditions under which they remain valid. The protocol implication is that AI-enabled reports and control decisions require explicit semantics beyond the traditional definition of measurement quantities and resource configurations. This issue is not limited to CSI. Beam management, reference signal adaptation, and predictive mobility control all raise a similar question: when an AI output affects a decision at the peer node, what must be standardized so that the decision remains interoperable and reliable?

A practical AI-native air interface cannot be standardized by adding a generic "AI mode" to the radio protocol. Such an approach would be too coarse to support meaningful interoperability. AI may be used to assist various radio functions, and each of these functions may have different assumptions, timescales, reliability requirements, and fallback needs. At the same time, 3GPP should avoid standardizing the internal form of AI itself. Model architectures, training datasets, optimization methods, inference engines, and model weights should remain implementation specific because they evolve rapidly and are central to vendor differentiation. The appropriate standardization target lies between these two extremes [11]. 6G should define the protocol-visible objects that describe how AI-enabled radio functions are exposed, configured, interpreted, monitored, and safely controlled. These objects play a role similar to existing radio objects, e.g., a CSI report configuration does not specify how a UE estimates the channel, but it specifies what report is expected, when it is provided, and how the BS should interpret it. Similarly, an AI-native task profile should not specify how a model is trained, but it should define the semantics of the AI-enabled function and the conditions under which its output can be used.

| Main concept | Existing 3GPP basis | Proposed position | Signaling and ownership |
|---|---|---|---|
| AI task profile | Rel-19 defines feature-specific AI functionalities and inference configurations. | **Extension:** a common abstraction describing an AI-assisted radio function, its operating conditions, and peer-visible behavior. | Configured by the BS through RRC. |
| Capability description | Rel-19 supports feature-specific UE AI capability reporting. | **Extension:** indicates supported profiles, versions, operating conditions, processing constraints, and reporting support. | UE → BS through RRC capability signaling. |
| Profile identifier | Rel-19 uses functionality, inference configuration, associated identifier. | **Extension:** profileId and profileVersion identify peer-visible radio behavior, not necessarily a proprietary model. | Configured by RRC and referenced by MAC CE or DCI when needed. |
| Input/output description | Rel-19 defines feature-specific measurement resources and report formats. | **Extension:** standardizes input resource references and output semantics while leaving internal features implementation-specific. | BS configures inputs; UE reports configured outputs. |
| Validity indication | Rel-19 supports configuration-specific applicable/inapplicable reporting. | **Extension:** indicates whether a named profile remains usable within its configured scope and prediction horizon. | Reported by the UE or evaluated by the BS. |
| Confidence report | No generic cross-vendor confidence semantics are defined. | **Proposed 6G direction:** an optional, quantized and calibrated indication tied to a defined event, observation window, prediction horizon, and network action. | UE → BS when configured; otherwise network-internal. |
| Activation and selection | Feature-specific RRC, MAC CE, and DCI mechanisms already exist. | **Extension:** selects only among preconfigured, deterministically interpretable profiles. | The BS controls activation and per-transmission selection. |
| Fallback procedure | Existing procedures support deactivation, release, and conventional operation. | **Extension:** explicitly defines the baseline profile, triggers, effective time, and re-entry hysteresis. | The UE may request fallback; the BS selects the transmitted profile. |
| OAM model lifecycle and pairing | TS 28.105 specifies model training, testing, deployment, monitoring, updating, and withdrawal. | **Extension:** a pairing identifier is needed only when cross-node model compatibility is required. | OAM owns model availability; the BS owns access stratum activation; the UE owns local model availability and status reporting. |

**Table 1: Relationship between existing 3GPP mechanisms and proposed candidate 6G protocol objects.**

Table 1 compares existing 3GPP mechanisms and candidate 6G protocol objects in the proposed framework. Common conventions apply across the candidate protocol objects in Table 1. The functionality identifier defines the namespace, while a profile identifier is unique within that functionality and is paired with a profile version that changes when its interoperable behavior changes. An associated identifier retains its Release-19 meaning, whereas a model or pairing identifier is used only when cross-node compatibility requires it. Objects are scoped to the applicable UE, serving cell, and operating region; one or more profiles may be configured, with one fallback profile per adaptive profile. Candidate RRC objects would use bounded ASN.1 fields and be updated semi-statically and only when the functionality is configured. MAC CE, DCI, or UCI would reference preconfigured profiles through compact fields. Consequently, there is no per-grant overhead unless fast selection or reporting is enabled; the reduced-density demodulation reference (DMRS) example in Table 3 uses a one-bit DCI selector, a one-bit validity or fallback indication, and an optional two-bit confidence report.

The AI task profile is the semantic anchor for an AI-enabled radio function. It identifies the task, operating conditions, and meaning of its outputs without specifying the underlying model. A corresponding capability description indicates which profiles the UE supports and any constraints on bandwidth, rank, mobility range, processing timeline, reporting, or fallback. The BS can therefore configure a profile only when both nodes have compatible assumptions. Within a task, a profile identifier refers to a protocol-visible operating mode rather than a particular neural network. For example, CSI-prediction profiles may differ in prediction horizon, input measurement resources, or output representation. The identifier gives the UE and BS a compact, unambiguous reference to common external behavior while allowing vendors to use different proprietary models.

Each profile also identifies the input-resource associations and output semantics needed for interoperability. Raw measurements, local history, and internal features normally remain within the UE or BS; only shared observations or reported features need standardized associations. For an output, the profile defines its meaning, quantization, referenced time or resource, and intended use. These requirements apply to predicted channel states, compressed representations, validity indications, and confidence reports. The profile also states whether the output describes the current state or predicts a future state and defines the horizon after which it is stale. Thus, the protocol standardizes only the information necessary for the peer to interpret and safely use the output, rather than imposing a universal representation across AI tasks.

AI-native operation requires explicit and interoperable meanings for validity and confidence. Validity can be a binary, configuration-specific indication of whether a named profile can be used within its configured operating region and validity horizon; it corresponds conceptually to configuration-specific applicability reporting in 3GPP Release 19. Confidence quantifies the likelihood that the profile output will meet a defined performance event and should not be interpreted as an uncalibrated, vendor-specific model score. Each profile therefore defines the reference event, observation window, prediction horizon, confidence quantization, statistical error or coverage interpretation, decision thresholds, and associated network actions. The standard need not prescribe how the underlying confidence is calculated, but implementations using the same confidence indication must satisfy the same externally testable statistical meaning.

Finally, every AI-native profile that can affect air interface behavior should include monitoring and fallback semantics. Monitoring determines whether the profile remains suitable for use after activation. Fallback defines the conservative behavior when the AI-enabled mode becomes unreliable or invalid. This is particularly important for link-critical functions such as reference signal adaptation. A reduced-density reference signal mode, for instance, should always be associated with a baseline reference signal configuration to which the system can return. Fallback is the mechanism that allows AI-native operation to be introduced safely.

## III. CANDIDATE PROTOCOL AND PROCEDURE

Cellular systems already have a mature layering structure that separates long-term configuration, medium-timescale control, fast physical layer operation, and network management. A practical AI-native design should build on this structure rather than bypass it. The purpose of the protocol architecture is therefore to determine where AI-related semantics should reside within the existing RRC, MAC, physical layer, and management framework. This layered view is important because AI-native functions operate across different timescales. Some aspects evolve slowly, such as UE capability, supported AI task profiles, model availability, and operator policy. Other aspects change at medium timescales, such as whether an AI profile is active, suspended, under validation, or in fallback. Still others must be handled close to the physical layer, e.g., measurement reporting may occur within a few slots. A clean architecture should place each AI-related function at the layer whose timing and reliability properties match the function being controlled, as illustrated in Figure 1(a).

RRC is the natural layer for defining the semantic framework of AI-native operation. In the proposed framework, RRC would configure the AI task profile, the radio scope to which it applies, the input and output representations, the associated measurement resources, the expected timing behavior, the validity assumptions, the confidence reporting format, and the fallback profile. In other words, this candidate RRC configuration would establish what an AI-enabled function means before lower layers are allowed to use it dynamically. RRC is also the right place for capability discovery. AI capability cannot be represented by a single feature bit because AI support is often conditional. The role of RRC capability signaling is to expose these constraints in a structured way so that the network can configure AI-native operation only when the UE and BS share compatible assumptions. This preserves the multi-vendor nature of cellular systems: each vendor can implement the AI function differently, but the protocol-visible capability and configuration semantics remain common.

Although RRC defines the semantic envelope, it is too slow for frequent adaptation. AI-native functions may need to be activated, suspended, resumed, or switched as radio conditions change. This is the role of MAC. In the proposed framework, MAC would manage the operational state of already configured AI profiles. For example, an AI-assisted CSI profile may be configured by RRC but activated only when the channel has been sufficiently observed and the UE reports adequate confidence. A candidate MAC procedure could support such transitions without requiring full RRC reconfiguration, allowing AI-native operation to adapt at a timescale that is practical for radio control. The MAC layer is also well suited to fallback management. When an AI-enabled mode becomes unreliable, the system should be able to return to a conservative operating mode that is already configured and understood by both sides. The candidate MAC procedure could coordinate this transition by activating a fallback profile, suspending the AI-assisted profile, or switching to a less aggressive configuration.

The physical layer is where AI-native operation becomes directly tied to signals, measurements, and scheduling-time decisions. AI-enabled functions ultimately rely on physical observations and often influence physical layer behavior. The proposed framework would use physical-layer mechanisms to provide the reference signal, measurement, and resource associations needed by the AI task profiles configured at higher layers. Candidate RRC signaling could configure several profiles, a candidate MAC CE could activate a subset, and candidate DCI signaling could select the profile applicable to a transmission or reporting occasion. Similarly, candidate UCI signaling could carry compact indications such as validity flag or fallback request. In short, dynamic layer 1 signaling should carry only what is needed for immediate radio operation: profile selection, triggering, compact reporting, and urgent reliability indications.

| Responsibility | UE-side model | Network-side model | Collaborative/two-sided model |
|---|---|---|---|
| Input ownership | UE-local measurements, history, and model state; BS configures relevant resources. | Network observations and standardized UE measurement reports. | Inputs are split; each node retains local data unless a shared representation is configured. |
| Data exposed across air interface | Configured outputs, measurements, validity, confidence, or performance metrics; raw inputs and internal features remain local. | UE provides only configured measurements or reports; model state and outputs remain network-internal. | Only specified measurements, representations, status, and, when needed, pairing/version information. |
| Inference timing | Within the UE reporting or receiver processing deadline. | Within the BS scheduling or radio resource management deadline. | Coordinated within a configured horizon, effective time, and radio control deadline. |
| Validity, confidence, and metric calculation | UE determines validity and, if configured, confidence or a performance metric. | BS determines validity/confidence and calculates the performance metric. | Each node assesses local validity/confidence; the BS combines UE status with network observations for end-to-end evaluation. |
| Monitoring, management, and decision authority | Network initiates monitoring and makes access stratum management decisions. UE manages local model availability; BS controls activation and fallback. | OAM manages the model lifecycle; the BS performs inference, monitoring, activation, and fallback. | Each domain manages its local component. The BS controls peer-visible behavior and fallback; the UE may report invalidity or request fallback. |

**Table 2: Responsibility allocation for different model placements.**

3GPP TS 28.105 already specifies management functions for the AI lifecycle, including training, testing, inference emulation, deployment, inference, performance monitoring, update, and withdrawal [8]. These operations, administration, and maintenance (OAM) lifecycle states are distinct from access stratum functionality configuration, activation, and applicability [12]. OAM owns the operational availability, version, update, and withdrawal of network-side models and associated resources; the BS access stratum owns radio profile configuration, activation, scheduling-time selection, monitoring, and fallback; and the UE owns local model availability and reports the applicability of configured UE-side functionality. A model may therefore be operationally available at the OAM level while its associated radio profile remains configured but inactive. If OAM withdraws a model or an update fails, the BS stops new activations and selects the configured baseline profile. If a UE-side model becomes unavailable, the UE reports the affected functionality as inapplicable and the BS deactivates or releases the profile. Table 2 summarizes the allocation of responsibilities for UE-side, network-side, and collaborative models.

Figure 1(b) illustrates the proposed operation procedure. After capability discovery establishes compatible task support and operating constraints, the BS uses RRC to configure one or more AI task profiles, including their scope, resource associations, output semantics, validity and confidence reporting, and fallback behavior. Configuration establishes a shared interpretation but does not imply immediate activation. A profile may first operate in an observation state while its validity is assessed using current measurements and model availability. The allocation of inference, monitoring, metric calculation, and management responsibilities depends on model placement, as summarized in Table 2. Regardless of placement, the BS retains decision authority whenever activation changes peer-visible radio behavior; the UE reports its local status and may request fallback when configured.

Once a profile is activated, the semantics configured by RRC allow fast control to remain compact. Candidate MAC CE, DCI, and UCI signaling can activate or select preconfigured profiles and convey configured validity, confidence, or fallback indications. Monitoring combines conventional radio outcomes, such as hybrid automatic repeat request (HARQ) feedback and CSI, with profile-specific indications. If reliability degrades, the BS may first select a more conservative profile or shorter prediction horizon; if the profile becomes invalid or degradation is severe, it selects the configured baseline from a defined effective time. The UE may request fallback but does not autonomously reinterpret BS-transmitted resources. Re-entry requires renewed validity and satisfaction of the configured confidence threshold over a validation window with hysteresis.

Table 3 instantiates the proposed procedure for reduced-density DMRS that may be used by a neural receiver in the downlink from UE capability and RRC configuration of baseline and reduced-density profiles through activation, per-grant selection, monitoring, fallback, and re-entry. For the reduced-density DMRS profile $P_1$, let $E_1$ denote successful first-transmission decoding of a physical downlink shared channel (PDSCH) transport block using $P_1$ within the configured processing deadline. Based on observations collected over the preceding window, the confidence quantity $q_1$ is a calibrated estimate on the probability of $E_1$ over the next prediction window. Each implementation maps its internal score to the standardized confidence classes using held-out empirical calibration data independent of model training. High confidence means $q_1 \geq q_{\text{use}}$ and permits continued use of $P_1$.

| Step | Signaling and illustrative candidate fields | Function, timing, and action |
|---|---|---|
| 1. Capability | UE → BS capability signaling: supported DMRS profile versions and pattern families; supported bandwidth/subcarrier spacing/rank/mobility region; validity and confidence reporting; fallback-request support | Establish whether the UE supports the reduced-density DMRS operation and its required reporting procedures |
| 2. RRC configuration | BS → UE RRC: functionalityId, profileId, profileVersion, dmrsPatternId, operatingRegion, validityHorizon, confidenceThresholds=$\{q_{\text{use}}, q_{\text{fb}}\}$, confidenceReportConfig, fallbackProfileId=$P_0$, $N_{\text{down}}$, and $N_{\text{up}}$ | Preconfigure baseline profile $P_0$ and reduced-density profile $P_1$, including validity, confidence, fallback, and re-entry semantics |
| 3. Validity reporting | UE → BS: profileId, profileVersion, validityFlag, and validity horizon; event-driven RRC or candidate fast UCI when configured | validityFlag=true indicates that $P_1$ is currently usable within its configured operating region; false causes the BS to select $P_0$ |
| 4. MAC-CE activation | BS → UE candidate MAC CE: profileId, activationState, and effectiveSlot | Activate $P_1$ among the configured profiles from an unambiguous effective slot |
| 5. DCI selection | BS → UE candidate DCI: one-bit dmrsProfileIndicator | Select $P_0$ or $P_1$ for each PDSCH grant; only preconfigured, deterministically interpretable patterns may be selected |
| 6. Monitoring and confidence | HARQ feedback plus optional UE → BS UCI**:** two-bit confidenceClass and one-bit fallbackRequest | 11: high - continue $P_1$; 10: medium – use conservative scheduling or verification; 01: low – select $P_0$; 00: unavailable – do not rely on $P_1$ |
| 7. Fallback | BS selects $P_0$ through the next applicable DCI grant or a configured MAC-CE effective slot. Trigger: validityFlag=false, one low/unavailable confidence report, $N_{\text{down}}$ consecutive medium-confidence reports, or a BS-side monitoring failure | Provide deterministic recovery. The UE may request fallback but does not autonomously reinterpret the BS-transmitted DMRS pattern |
| 8. Re-entry | validityFlag=true and $N_{\text{up}}$ consecutive high-confidence reports, followed by BS activation/selection of $P_1$ | Confirm renewed usability and prevent oscillation between $P_0$ and $P_1$ |

**Table 3: Illustrative end-to-end candidate procedure for a reduced-density PDSCH DMRS profile. The field names and bit widths are illustrative candidate 6G design examples and are not existing or agreed 3GPP information elements.**

Medium means $q_{\text{fb}} \leq q_1 < q_{\text{use}}$ and triggers conservative scheduling or verification. Low means $q_1 < q_{\text{fb}}$, while unavailable means that no calibrated estimate can be provided; either condition causes the BS to select the baseline profile $P_0$. An indication of false validity or one low-confidence report triggers fallback, while $N_{\text{down}}$ consecutive medium-confidence reports may also trigger fallback. Re-entry requires validity to become true and confidence to remain high for $N_{\text{up}}$ consecutive reports.

Protocol-format conformance and AI-function performance conformance should be evaluated separately. Protocol-format conformance verifies encoding of the validity and confidence fields, association with the correct profile and version, reporting time, and execution of the specified network action. AI-function performance conformance evaluates the statistical meaning of confidence using common black-box tests across vendors. The test distributions specify channel models, delay spreads, signal-to-noise ratio (SNR) ranges, modulation and coding scheme (MCS) ranges, ranks, Doppler conditions, minimum sample counts, and independent random seeds within the configured operating region. For each confidence class other than low or unavailable, the estimate on the observed probability of $E_1$ must meet the requirement advertised for that class. In specified out-of-distribution conditions, the implementation must report validity as false or confidence as low or unavailable within a time limit, and the rate of erroneous high-confidence reports must remain below a profile-specified limit.

The following case study situates this reduced-density DMRS profile within a broader progression toward increasingly AI-dependent reference signal designs.

## IV. CASE STUDY: NEURAL RECEIVER ASSISTED REFERENCE SIGNAL ADAPTATION

Neural receivers provide a useful case study for illustrating the protocol framework developed in the previous sections because they sit at the boundary between implementation freedom and air interface coordination. If a UE uses a neural network only to improve channel estimation, equalization, or demapping while receiving the same reference signals and following the same control procedures as a conventional UE, then the neural receiver remains an implementation detail [13]. The BS does not need to know how the receiver is designed, trained, or executed. Interoperability is preserved because the standardized air interface behavior is unchanged.

The standardization question becomes more interesting when the receiver capability begins to affect what the BS transmits.

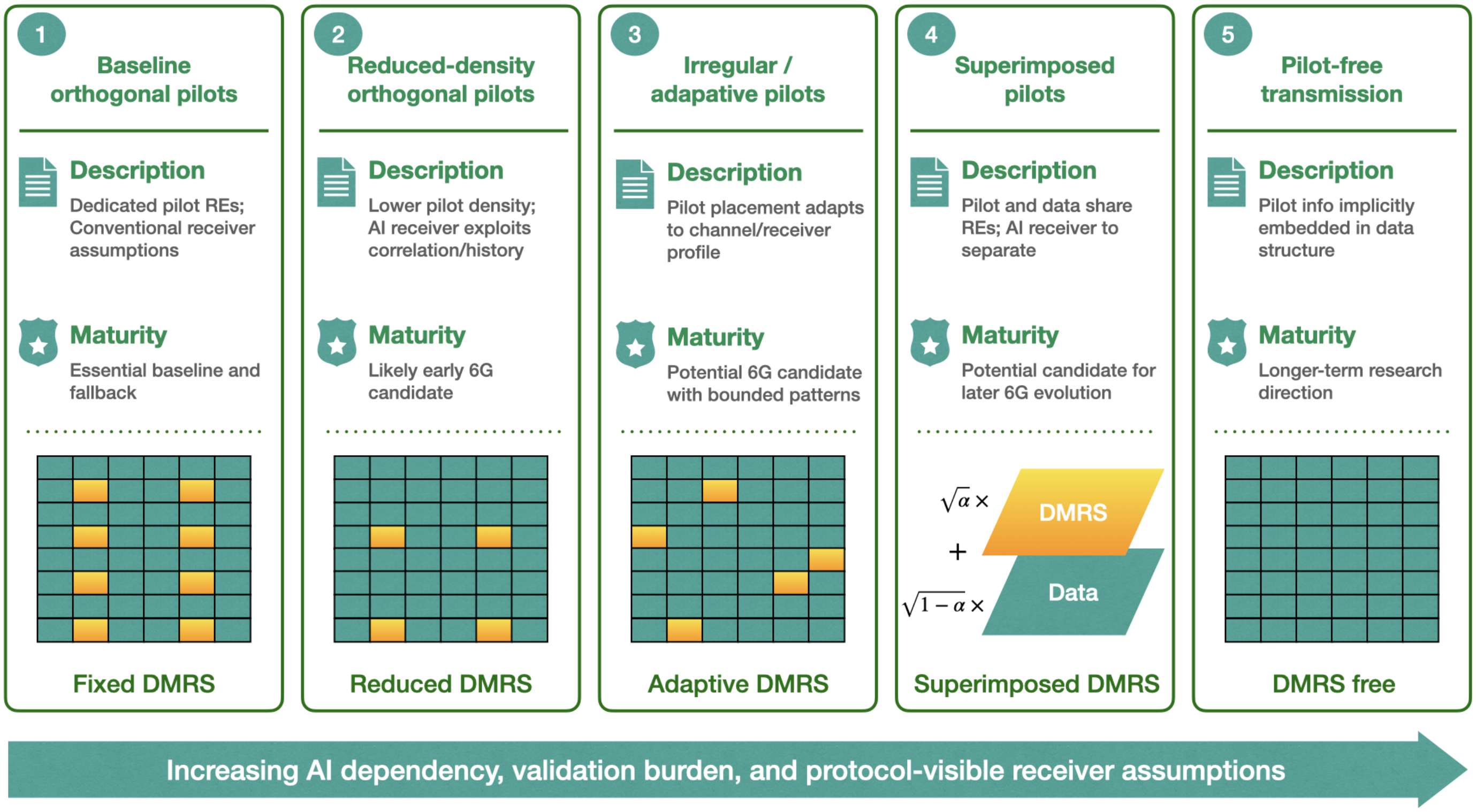

**Figure 2: An illustration of different levels of neural receiver assisted reference signal adaptation.**

A neural receiver may be able to exploit time-frequency correlation, spatial structure, channel history, or learned impairment characteristics more effectively than a conventional receiver. This may allow the system to reduce DMRS overhead, adapt pilot placement, or use more aggressive reference signal designs under suitable conditions. At that point, the receiver is no longer only a proprietary implementation behind a fixed interface. The BS must know that the UE can support a particular reference signal configuration, when that support is valid, how reliable the UE considers the receiver operation, and how the system should fall back if the receiver becomes unreliable. A useful way to understand the standards impact is to view neural receiver assisted DMRS operation as a continuum, as illustrated in Figure 2.

**1) Baseline Orthogonal Pilots:** At one end is the baseline orthogonal DMRS design, where pilot resource elements (REs) are dedicated to the reference signal and the receiver follows conventional assumptions. In this regime, a neural receiver can still provide implementation gain, but the standards impact is minimal because the transmitted signal and the protocol interpretation of DMRS remain unchanged. This mode is essential because it provides the conservative anchor for interoperability and fallback. Even if more advanced AI-assisted modes are introduced, the baseline DMRS profile gives the UE and BS a robust configuration to return to when the AI-assisted mode becomes invalid or insufficiently reliable.

**2) Reduced-density Orthogonal Pilots:** A moderate step beyond the baseline is reduced-density orthogonal DMRS. Here, the pilot density is reduced, but pilots still occupy dedicated orthogonal REs. The neural receiver is expected to recover the channel from fewer observations by exploiting correlation over time, frequency, antennas, or recent channel history. This changes the standards problem in a manageable way. The air interface can still define a bounded set of DMRS profiles, and the BS can select a reduced-density profile only when the UE has declared support for the corresponding operating envelope. The AI model remains proprietary, but the reduced-DMRS capability, validity assumptions, and fallback profile become protocol visible. Such a design is attractive for early 6G consideration because it preserves the familiar interpretation of DMRS while allowing AI to reduce overhead.

**3) Irregular/adaptive Pilots:** Irregular or adaptive DMRS represents a deeper level of AI-native operation. In this case, the pilot pattern is not simply a lower-density version of a fixed grid. Instead, pilot placement may adapt to channel conditions, mobility, receiver class, cell-specific environment, or other contextual information. A neural receiver may make such patterns viable by learning to interpolate from non-uniform observations and by exploiting structure that is difficult to capture with conventional interpolation. However, the standards impact also increases. The UE and BS must have a common understanding of the allowed pattern family, the profile identifier, the conditions under which the pattern is valid, and the mechanism by which the pattern is selected. The standard does not need to prescribe how the neural receiver performs interpolation, but it does need to define the protocol semantics of the adaptive DMRS profile.

**4) Superimposed Pilots:** A larger qualitative shift occurs with superimposed pilot transmission [14]. In conventional transmission, pilot and data symbols are separated in REs. In superimposed transmission, pilot and data components share the same time-frequency resources through a controlled embedding rule, for example, through a power split between pilot and data components. This design can potentially reduce explicit pilot overhead because pilot information is embedded together with data. However, it also changes the nature of the air interface. The receiver must jointly interpret the pilot and data components, while the transmitter must embed them according to a structure that the receiver has been trained to understand. This is why superimposed pilot transmission is an example of two-sided training. A one-sided neural receiver can be introduced without changing the transmitted waveform, but superimposed pilots require joint assumptions at both ends of the link. The transmitter-side embedding rule and receiver-side neural processing are coupled. If the BS changes the superposition rule without a compatible receiver assumption, the UE may no longer be able to disentangle pilot and data reliably. Conversely, a neural receiver trained for one superposition profile may not generalize to a different profile with a different mobility regime or channel condition. Therefore, the relevant standardization object is not the trained model, but the superposition profile and its protocol envelope.

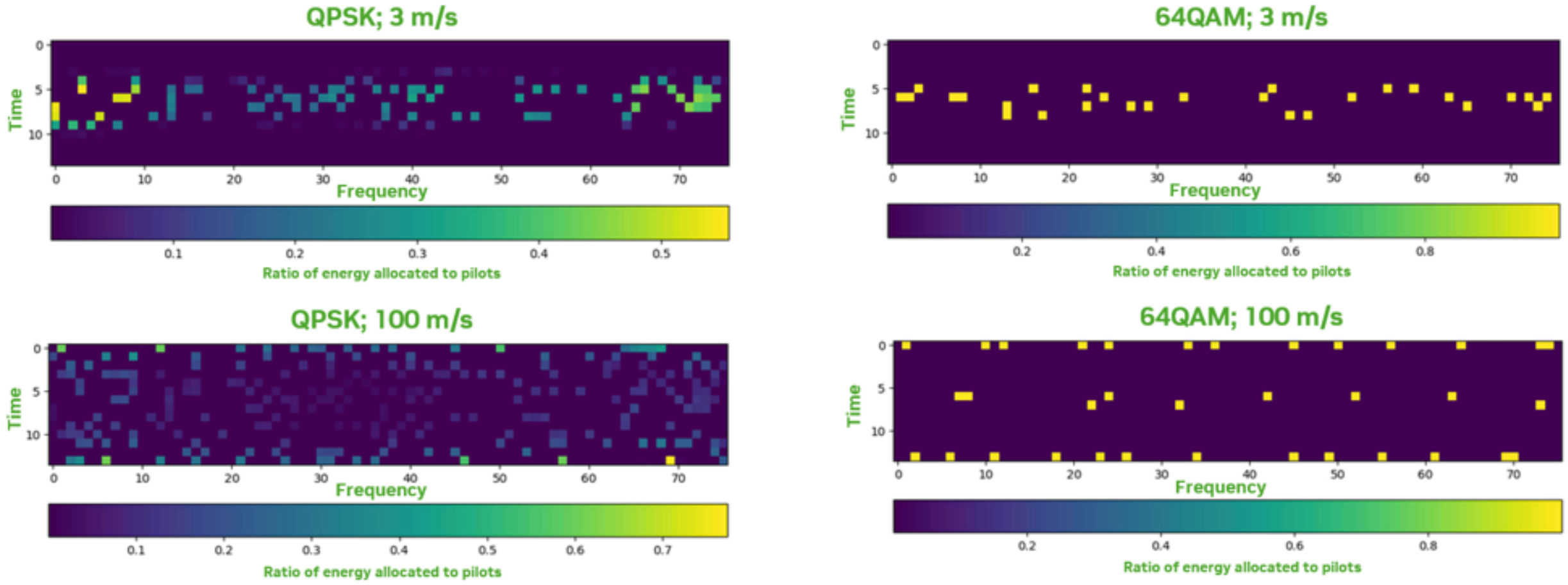

**Figure 3: Learned pilot-to-data power splits for superimposed pilot transmission in two-sided training.**

Figure 3 shows representative learned pilot-energy allocations obtained by leveraging the publicly available Sionna superimposed-pilots tutorial [15]. The baseline configuration uses 14 orthogonal frequency-division multiplexing (OFDM) symbols, 76 subcarriers, 30 kHz subcarrier spacing, and a clustered delay line (CDL) channel with 100 ns delay spread. The receiver comprises four 128-channel residual convolutional blocks. The receiver weights and pilot-energy allocation are jointly optimized by maximizing the bit-metric decoding rate. As shown in Figure 3, under low mobility, the model may concentrate pilot energy on a smaller set of resources because the receiver can exploit stronger temporal and frequency correlation. Under higher mobility, the allocation becomes more dispersed, reflecting the need to refresh channel information more frequently. Modulation order also matters. For higher modulation, where residual channel estimation error can have a stronger impact on detection, the learned allocation tends to preserve more pronounced pilot-bearing components than in lower-order modulation cases. From a protocol perspective, superimposed pilots therefore require richer support than reduced-density orthogonal pilots. A candidate 6G design could use RRC to configure the superposition profile, including the pilot embedding rule, permitted power-split range, validity assumptions, confidence reporting format, and fallback DMRS profile. A candidate MAC procedure could activate or suspend the profile based on receiver readiness and radio conditions. Candidate physical layer signaling could select among configured profiles or trigger compact validity reporting. During operation, the UE may report whether the current profile remains reliable, while the BS may respond by adjusting scheduling, reverting to orthogonal DMRS, or selecting a less aggressive profile.

**5) Pilot-free Transmission:** At the far end of the continuum lies pilot-free transmission, where pilot information is implicitly embedded in the data structure and the receiver relies on learned inference rather than explicit pilots. This is the most ambitious point in the design space and also the most challenging from a standards perspective. It would require strong transmitter-receiver co-design, extensive validation, carefully defined conformance behavior, and robust fallback. For this reason, pilot-free transmission may be better viewed as a longer-term research direction than as an immediate 6G standardization target from day 1. It remains valuable, however, because it clarifies the direction of progress: as AI becomes more deeply integrated into the physical layer, receiver assumptions become increasingly protocol visible.

## V. CONCLUSION AND FUTURE OUTLOOK

AI-native 6G requires common protocol semantics when AI affects peer-visible radio behavior. Building on 3GPP Release 19 mechanisms, this article proposes a candidate 6G protocol framework for describing AI-enabled functions, configuring and activating profiles, interpreting validity and confidence, monitoring performance, and providing deterministic fallback. Model architectures, training methods, and weights can generally remain implementation specific, while compatible representations or paired behavior may need to be identified when transmitter and receiver operation is coupled. The reduced-density DMRS and superimposed-pilot examples illustrate how increasingly AI-dependent physical layer operation creates progressively stronger interoperability requirements. Evolution toward AI-native 6G is therefore likely to be gradual, beginning with bounded profiles, explicit operating regions, testable confidence semantics, and conservative fallback. The objective is not to standardize intelligence itself, but the protocol framework through which AI-assisted radio behavior can operate reliably across vendors and deployments.